\documentclass[%
 reprint,
superscriptaddress,
 amsmath,amssymb,
 aps,
]{revtex4-2}

\usepackage{graphicx}
\usepackage{dcolumn}
\usepackage{bm}
\usepackage{gensymb}
\usepackage{units}

\begin{document}

\preprint{APS/123-QED}

\title{Quantum and thermal noise in coupled non-Hermitian waveguide systems with different models of gain and loss}

\author{Osmery Hernández}
\affiliation{%
 Department of Electrical, Electronic and Communications Engineering,Public University of Navarra, 31006 Pamplona, Spain
 }%
\author{Iñigo Liberal}
\thanks{Corresponding author: inigo.liberal@unavarra.es}%
\affiliation{%
 Department of Electrical, Electronic and Communications Engineering,Public University of Navarra, 31006 Pamplona, Spain
 }%
 \affiliation{
 Institute of Smart Cities, Public University of Navarra, 31006 Pamplona, Spain
}%

\date{\today}

\begin{abstract}
Non-Hermitian (NH) photonic systems leverage gain and loss to open new directions for nanophotonic technologies. However, the quantum and thermal noise intrinsically associated with gain/loss affects the eigenvalue/eigenvector structure of NH systems, as well as its practical noise performance. Here, we present a comparative analysis of the impact of different gain and loss mechanisms on the noise generated in gain-loss compensated NH waveguide systems. Our results highlight important differences in the eigenvalue/eigenvector
structure, noise power, photon statistics and squeezing. At the same time, we identify some universal properties such as gain-loss compensation, broken to unbroken phase transitions, coalesce of pairs of eigenvectors, and linear scaling of the noise with the length of the waveguide. We believe that these results provide a more global understanding
on the impact of the gain/loss mechanism on the noise generated in NH systems.
\end{abstract}

\maketitle

\section{Introduction}

Non-Hermitian (NH) systems are ubiquitous among real physical systems since describing a system as an isolated entity is often impossible or inaccurate \cite{breuer2002theory}. Since any external environment can either pump into or retract energy from a system, NH systems are usually associated with gain, loss, or a combination of both \cite{wang2023non}. In addition, Hermitian operators have real eigenvalues, while the eigenvalues of NH operators are generally complex numbers. However, a reduced group of NH operators satisfying a weaker constraint than Hermiticity can still have a real spectrum in a given region in parameter space. These systems exhibit PT-symmetry \cite{bender1998real,bender2007making}, or in other words, their Hamiltonian ($\boldsymbol{\widehat{H}}$) commutes with the joint operation of Parity ($\boldsymbol{\widehat{P}}$) and Time Reversal ($\boldsymbol{\widehat{T}}$).

Photonics have been an instrumental platform in the demonstration of PT-symmetric systems, given the equivalence between the Schrödinger equation and the paraxial wave approximation of Helmholtz equation, where the complex refractive index plays the role of the complex potential in the NH Hamiltonian
\cite{miri2019exceptional,wang2023quantum,zhao2018parity}. Therefore, by engineering the refractive index through the appropriate combination of gain and loss, it is possible to create the real even and imaginary odd refractive index profiles characteristic of PT-symmetric systems. The transition between the regions where PT-symmetry is conserved, or unbroken phase, and the broken phase usually feature spectral degeneracies known as Exceptional Points (EPs) \cite{feng2017non,zhao2018parity}. At these transition points, the eigenvalues coalesce, and so do the eigenvectors; thus the eigenspace is skewed. EP degeneracies have also been studied in the context of Liouvillian operators, scattering matrix approaches or using coupled mode theory \cite{wang2023non}.

PT-symmetry and EP degeneracies endow physical systems with intriguing non-trivial properties inaccessible to Hermitian systems, thus motivating intense research on the field in the last few years \cite{wang2023non,ozdemir2019parity,parto2020non,el2018non,feng2017non,miri2019exceptional,li2023exceptional}. Examples of such counterintuitive effects include unidirectional reflectionless light propagation, and therefore, unidirectional invisibility \cite{lin2011unidirectional,feng2013experimental,huang2017unidirectional}, non-reciprocal light propagation \cite{ruter2010observation,chang2014parity}, loss-induced transparency \cite{guo2009observation,tschernig2022branching} and lasing \cite{peng2014loss}, PT-symmetric lasers \cite{hodaei2014parity,feng2014single} and CPA-lasers \cite{wong2016lasing,longhi2010pt}, and chiral mode switching \cite{doppler2016dynamically}, to name a few. EPs have also been proposed to enhance the performance of sensors \cite{zhang2019quantum}. Although the noise performance of EP sensors is the subject of current debate \cite{loughlin2024exceptional,roy2021nondissipative}.

In the quantum optics realm, systems featuring PT-symmetry and EPs have been reported to influence quantum interference
\cite{klauck2019observation,zhou2022characterization,longhi2020quantum}, entanglement \cite{chakraborty2019delayed,antonosyan2018photon}, and decoherence \cite{gardas2016pt,dey2019controlling}. At the same time, it has been claimed that obtaining PT symmetry in quantum photonics systems combining loss and gain is not possible \cite{scheel2018symmetric} due to the additional noise contribution of gain at the quantum level, which changes the underlying eigenvalue structure. However, this study only considered models for linear gain and loss mechanisms. Different alternatives have been explored in studies on quantum PT-symmetry and EPs trying to overcome the noise issues associated to linear or phase-insensitive amplification, most of them either rely on passive implementations or consider a non-Hermitian subsystem embedded in a large Hermitian system \cite{wang2023quantum,klauck2019observation}. Recently, the potential of phase-sensitive amplification and deamplification have been showcased \cite{roy2021nondissipative,wang2023quantum}, demonstrating quadrature-PT symmetry and squeezing, what suggests that the model of gain and losses employed plays a critical role.

In this work, we present a general study on how the different quantum models of gain and loss impact the performance of Non-Hermitian photonic systems, and affect their underlying eigenvalue/eigenvector structure. To this end, we propose a theoretical framework to compute the quantum eigenmodes of the spatial evolution of coupled waveguides,
which applies to different models of gain and loss, as well as all their possible combinations. We also clarify how such algebraic quantum eigenmodes can be measured with conventional measurement setups at the output of the waveguides. Our results provide a global perspective into how the nature of gain and loss mechanisms define the nature
of Non-Hermitian phase transitions, including their eigenvalue/eigenvector structure, the potential existence of EPs, the unusal scaling of quantum noise generation with the length of the waveguide, and the generation of squeezing in non-Hermitian systems. Our results also contribute to understanding the noise generated in gain-loss compensation systems,
as a function of the models of gain and loss.

\section{Theoretical framework}

In this section we introduce a theoretical framework to model how the presence of gain and loss and their linear or parametric nature influence the evolution of quantum light states in photonic systems characterized by modes propagating along a given distance while exchanging excitations between them. For instance, that might be the case of
a finite number of coupled waveguides or resonators. This framework not only provides an analytical solution to the evolution equation of the coupled modes, allowing for the computation of photon statistics, but also addresses intuitively the question of how to design the measurement setup to characterized the singular eigenvector/eigenvalue properties of non-Hermitian systems.


\begin{figure*}
\includegraphics[width=0.8\textwidth]{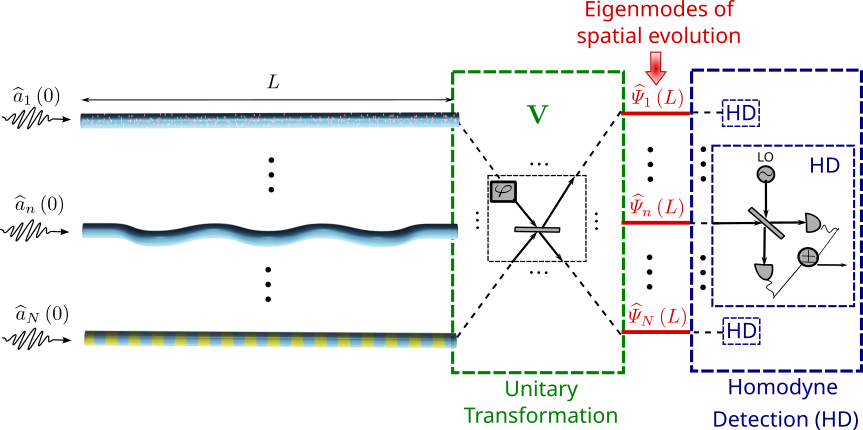}
\caption{\label{fig:General_theory}Shematic depiction of $N$ interacting photonic waveguides with different classes of gain and loss. At the output of the waveguides, a unitary transformation induces a change of basis of the photonic modes. Finally, homodyne detection enables the characterization of the photon statistics of the eigenoperators
of the waveguide system.}
\end{figure*}

\subsection{Spatial evolution in coupled non-Hermitian waveguides}

As schematically depicted in Fig.~(\ref{fig:General_theory}), let us start by mathematically modeling a photonic system where a finite set of photonic quantum modes evolve along a given distance and might couple with each other while propagating. Provided that the systems considered are in general non-Hermitian, they might exhibit gain and losses, either linear or parametric. The propagation or spatial evolution of the modes in these photonic systems can be described through the differential equation
\begin{equation}
\partial_{z}\widehat{\mathbf{a}}\left(z\right)=\mathbf{M}\,\widehat{\mathbf{a}}\left(z\right)+\widehat{\mathbf{F}}\left(z\right)
\label{eq:dif_evolution}
\end{equation}

\noindent where $\widehat{\mathbf{a}}\left(z\right)=\left[\widehat{a}_{1}\left(z\right),\ldots,\widehat{a}_{N}\left(z\right),\widehat{a}_{1}^{\dagger}\left(z\right),\ldots,\widehat{a}_{N}^{\dagger}\left(z\right)\right]^{T}$ is the column vector of the photonic annihilation ($\widehat{a}_{m}\left(z\right)$) and creation ($\widehat{a}_{m}^{\dagger}\left(z\right)$) operators satisfying bosonic commutation relations ($\left[\widehat{a}_{m}\left(z\right),\widehat{a}_{n}\left(z'\right)\right]=0$
and $\left[\widehat{a}_{m}\left(z\right),\widehat{a}_{n}^{\dagger}\left(z'\right)\right]=\delta\left(z-z'\right)\,\delta_{mn}$), while $\widehat{\mathbf{F}}\left(z\right)=\left[\widehat{F}_{1}\left(z\right),\ldots,\widehat{F}_{N}\left(z\right),\widehat{F}_{1}^{\dagger}\left(z\right),\ldots,\widehat{F}_{N}^{\dagger}\left(z\right)\right]^{T}$ contains operators representing Langevin noise sources \cite{gardiner2004quantum,barnett1998quantum}, and $\mathbf{M}\in\mathbb{C}^{2N\times2N}$ is the spatial evolution matrix. It is important to point out that, according to our notation, $\mathbf{M}=-i\mathbf{H}$ with respect to the Hamiltonian description of NH waveguides, with the corresponding consequences in the physical interpretation of real and complex eigenvalues. 

The structure of the noise vector $\widehat{\mathbf{F}}\left(z\right)$ in (\ref{eq:dif_evolution}) will depend on the system under study: if there is gain or loss associated with the corresponding photonic mode and its linear or parametric nature. On the one hand, describing gain and losses through linear models involves the presence of thermal noise sources that can be represented through bosonic creation ($\widehat{f}_{n}^{\dagger}\left(z\right)$) or destruction ($\widehat{f}_{n}\left(z\right)$) noise operators,
respectively, with their associated commutator $\left[f_{m}\left(z\right),f_{n}^{\dagger}\left(z'\right)\right]=\delta\left(z-z'\right)\delta_{mn}$. Therefore, for a photonic mode $m$ experiencing linear losses $\alpha$, we can define the Langevin noise vector component $\widehat{F}_{m}$ in terms of a bosonic noise operator: $\widehat{F}_{m}\left(z\right)=\sqrt{2\alpha}\,\widehat{f}_{m}\left(z\right)$, with commutation relations $\left[\widehat{F}_{m}\left(z\right),\widehat{F}_{n}^{\dagger}\left(z'\right)\right]=2\alpha\,\delta\left(z-z'\right)\delta_{mn}$ \cite{haus2012electromagnetic,gardiner2004quantum}. Similarly, $\widehat{F}_{m}\left(z\right)=\sqrt{2g}\,\widehat{f}_{m}^{\dagger}\left(z\right)$ for a mode subject to amplification with linear gain $g$, and the associated commutator is given by $\left[\widehat{F}_{m}\left(z\right),\widehat{F}_{n}^{\dagger}\left(z'\right)\right]=-2g\,\delta\left(z-z'\right)\delta_{mn}$ \cite{haus2012electromagnetic,gardiner2004quantum}.

On the other hand, gain and loss can also arise from nonlinear or parametric phenomena leading to squeezing transformations and mixing creation and destruction photonic operators. Modeling both phenomena through the same process is not an arbitrary decision; it is justified because a parametric gain can also be regarded as a parametric loss mechanism, depending on the quadrature component we observe, as one of them is amplified while the other is attenuated given the phase-sensitive nature of the process \cite{gardiner2004quantum}. No additional noise sources are required for parametric gain/loss, and their behavior is fully contained within the matrix $\mathbf{M}$.

\subsection{Eigenoperators of spatial evolution}

In general, calculating photon statistics at the output of the waveguides can be a complicated task. However, the analysis simplifies for specific linear combinations of the waveguide modes for which their spatial evolution decouple. Importantly, such modes can be identified even for non-diagonalizable evolution matrices $\mathbf{M}$, as is typically the situation at the phase transition of NH systems, where eigenvectors coalesce. To show how this is the case, we introduce a new operator given by a linear combination of waveguide mode operators $\widehat{\varPsi}_{n}\left(z\right)=\sum\limits_{m=1}^{2N}c_{m}\widehat{a}_{m}\left(z\right)$. Following Eq.~(\ref{eq:dif_evolution}), the spatial evolution of $\widehat{\varPsi}_{n}\left(z\right)$ is given by 
\begin{multline}
\partial_{z}\widehat{\varPsi}_{n}=\sum\limits_{n=1}^{2N}\left(\sum\limits_{m=1}^{2N}c_{m}M_{mn}\right)\widehat{a}_{n}\left(z\right)+\sum\limits_{m=1}^{2N}c_{m}\widehat{F}_{m}\left(z\right)
\label{eq:linear_comb_spaceEvol_extended}
\end{multline}

\noindent From Eq.~(\ref{eq:linear_comb_spaceEvol_extended}) it is apparent that if such a linear combination corresponds to a left eigenvector of the dynamic matrix $\mathbf{M}$, i.e., $\sum\limits_{m=1}^{2N}c_{m}M_{mn} = \lambda_{n}c_{n}$, then
\begin{equation}
\partial_{z}\widehat{\varPsi}_{n}\left(z\right)=\lambda_{n}\widehat{\varPsi}_{n}\left(z\right)+\widehat{\zeta}_{n}\left(z\right)
\end{equation}

\noindent where $\widehat{\zeta}_{n}\left(z\right)=\sum\limits_{m=1}^{2N}c_{m}\widehat{F}_{m}\left(z\right)$, and we can clearly distinguish that the evolution of the eigenoperators defined through the left eigenvectors of the dynamic matrix $\mathbf{M}$ is not interacting, and can be computed independently. Accordingly, it is possible to obtain the eigenoperator at the output of the waveguide upon direct integration:
\begin{equation}
\widehat{\varPsi}_{n}\left(L\right)=\widehat{\varPsi}_{n}\left(0\right)\,e^{\lambda_{n}L}+e^{\lambda_{n}L}\int_{0}^{L}dz'e^{-\lambda_{n}z'}\widehat{\zeta}_{n}\left(z'\right)\label{eq:newModes_L}
\end{equation}

\noindent Equation (\ref{eq:newModes_L}) states that the eigenoperator at any given distance $L$ on the waveguide can be computed from the knowledge of the eigenoperator
at the start of the waveguide, i.e., at $z=0$ and its associated noise component.

The most appealing reading of this result is that for any given photonic quantum system, if we perform a unitary transformation that creates adequate linear combinations of the original bosonic modes, the associated evolution would be easily computed from the knowledge of the initial conditions in the system. We note that even if the matrix $\mathbf{M}$
is not diagonalizable, and their eigenvectors do not span the complete $\mathbb{C}^{2N}$ space, individual eigenmodes can nevertheless be physically separated with a unitary matrix $\mathbf{\mathbf{V}}$ that does span the complete $\mathbb{C}^{2N}$ space, and contains the eigenmode as a member of its basis. Therefore, the photon statistics of individual eigenoperators can be measured in practice even at the degeneracy points of NH systems. As previously pointed out, we must compute the left eigenvectors of the evolution matrix $\mathbf{M}$, or, equivalently, the right eigenvectors of the transposed evolution matrix $\mathbf{M}^{T}$. 

\subsection{Diagonalizable evolution matrices}

Naturally, the result showcased in Eq.~(\ref{eq:newModes_L}) can be more easily obtained for the particular case that the evolution matrix is diagonalizable, i.e. $\mathbf{M}=\mathbf{V^{-1}}\mathbf{D}\mathbf{V}$, where $\mathbf{V}$ is an invertible matrix whose columns are eigenvectors of the system and constitute a complete basis of $\mathbb{C}^{2N}$,
while the diagonal entries in the diagonal matrix $\mathbf{D}$ corresponds to the associated eigenvalues $\lambda_{n}$. In this case, one can simply change the basis of the photonic and noise operatores to $\widehat{\boldsymbol{\varPsi}}\left(z\right)=\mathbf{\mathbf{V}}\widehat{\mathbf{a}}\left(z\right)$ and $\widehat{\boldsymbol{\zeta}}\left(z\right)=\mathbf{\mathbf{V}}\widehat{\mathbf{F}}\left(z\right)$, respectively, so that their evolution can be described through a diagonal matrix
\begin{equation}
\partial_{z}\widehat{\boldsymbol{\varPsi}}\left(z\right)=\mathbf{D}\widehat{\boldsymbol{\varPsi}}\left(z\right)+\widehat{\boldsymbol{\zeta}}\left(z\right)
\label{eq:diag_evol}
\end{equation}

Therefore, the photon statistics of the eigenoperators associated with left eigenstates of the system can be easily computed by separating such eigenoperators through an optical network implementing the matrix $\mathbf{V}$ (see Fig.~\ref{fig:General_theory}). The crucial difference between diagonalizable and non-diagonalizable evolution matrices is that when $\mathbf{M}$ is non-diagonalizable, as it is common in NH systems, not all the light exiting the waveguides can be described via the eigenoperators, since the associated eigenvectors do not span the entire $\mathbb{C}^{2N}$ space. Despite this fact, the light and quantum noise associated with individual eigenoperators can be separated via unitary transformations, so that such algebraic singular eigenvectors/eigenvalues correspond with measurable photon statistics.

\subsection{Symmetries of the eigenvalues, eigenvectors and eigenoperators}

Next we discuss the symmetries of the eigenvectors of the matrix $\mathbf{M}$, which further clarify how to measure individual eigenoperators. To this end, let us start by defining $\boldsymbol{v}_{n}=\begin{bmatrix}\mathbf{x} & \mathbf{y^{*}}\end{bmatrix}^{T}$ as the eigenvectors of $\mathbf{M}^{T}$, i.e., $\mathbf{M}^{T}\boldsymbol{v}=\lambda\boldsymbol{v}$, so that the associated eigenoperator is
\begin{equation}
\widehat{\varPsi}_{n}\left(z\right)=\sum\limits_{m=1}^{N}x_{nm}\widehat{a}_{m}\left(z\right)+\sum\limits_{m=1}^{N}y_{nm}^{*}\widehat{a}_{m}^{\dagger}\left(z\right)
\end{equation}

\noindent and, analogously, the associated noise operator is given by
\begin{equation}
\widehat{\zeta}_{n}\left(z\right)=\sum\limits_{m=1}^{N}x_{nm}\widehat{F}_{m}\left(z\right)+\sum\limits_{m=1}^{N}y_{nm}^{*}\widehat{F}_{m}^{\dagger}\left(z\right)
\end{equation}

\noindent Additionally, given the structure of $\widehat{\mathbf{a}}\left(z\right)$ in (\ref{eq:dif_evolution}), it is always possible to write the evolution matrix $\mathbf{M}$ as a block matrix whose elements are the matrices $\mathbf{P}$, characterizing linear phenomena and coupling between modes, and $\mathbf{Q}$ accounting for nonlinear effects mixing annihilation and creation operators:
\begin{equation}
\mathbf{M}=\left[\begin{array}{cc}
\mathbf{P} & \mathbf{Q}\\
\mathbf{Q}^{*} & \mathbf{P}^{*}
\end{array}\right]
\label{eq:M}
\end{equation}

\noindent Therefore, the eigenvalue problem for the transpose of the evolution matrix reduces to the following conditions:
\begin{equation}
\mathbf{P}^{T}\mathbf{x}+\mathbf{Q}^{\dagger}\mathbf{y^{*}}=\lambda\mathbf{x}
\label{eq:rel1}
\end{equation}
\begin{equation}
\mathbf{Q}^{T}\mathbf{x}+\mathbf{P}^{\dagger}\mathbf{y^{*}}=\lambda\mathbf{y^{*}}
\label{eq:rel2}
\end{equation}

It is crucial to distinguish two cases in the structure of eigenvectors and, consequently, that of their associated operators, depending on whether the eigenvalues are real or complex numbers, i.e., depending on whether we are at the broken or unbroken phase of a NH system.

\subsubsection{Real eigenvalues}

If $\lambda\in\mathbb{R}$, by complex conjugating Eq.~(\ref{eq:rel2}), we obtain $\mathbf{Q}^{\dagger}\mathbf{x}^{*}+\mathbf{P}^{T}\mathbf{y}=\lambda\mathbf{y}$, and, by comparing with Eq.~(\ref{eq:rel1}) we can establish that $\mathbf{y}^{*}=\mathbf{x}^{*}$. Therefore, the eigenvectors associated with real eigenvalues exhibit the general structure $\boldsymbol{v}=\begin{bmatrix}\mathbf{x} & \mathbf{x}^{*}\end{bmatrix}^{T}$, leading to the eigenoperators:
\begin{equation}
\widehat{\varPsi}_{n}\left(z\right)=\widehat{\psi}_{n}(z)+\widehat{\psi}_{n}^{\dagger}(z)
\label{eq:Reigenv_HermitianOp}
\end{equation}

\noindent where $\widehat{\varPsi}_{n}\left(z\right)$ is explicitely written as a Hermitian operator with $\widehat{\psi}_{n}(z)=\sum\limits_{m=1}^{N}x_{nm}\widehat{a}_{m}(z)$
being a linear combination of physical waveguide mode operators. Then, the spatial evolution of this operator can be computed following Eq.~(\ref{eq:newModes_L}).

Furthermore, Eq.~(\ref{eq:Reigenv_HermitianOp}) reveals that the eigenoperators $\widehat{\varPsi}_{n}\left(z\right)$, associated with real eigenvalues, correspond to
a quadrature operators in the basis of the operators $\widehat{\psi}_{n}(z)$. Thus, this notation reveals that the photon statistics of such eigenoperator can be measured with conventional techniques consisting of: 1) a unitary transformation that changes the basis to another one in which a selected eigenmode of $\mathbf{M}^{T}$ is a member of the basis, and 2) homodyne detectors to extract the associated photon statistics, as eigenoperators of real eigenvalues represent quadrature components in this new basis (see Fig.~\ref{fig:General_theory}).

\subsubsection{Complex eigenvalues}

A different eigenvector/eigenvalue structure arises when $\lambda\in\mathbb{C}$. After complex conjugation of Eq.~(\ref{eq:rel1}) and Eq.~(\ref{eq:rel2}) it can be concluded that if $\lambda$ is an eigenvalue with eigenvector $\boldsymbol{v}=\begin{bmatrix}\mathbf{x} & \mathbf{y}^{*}\end{bmatrix}^{T}$, then $\boldsymbol{w}=\begin{bmatrix}\mathbf{y} & \mathbf{x}^{*}\end{bmatrix}^{T}$ is also an eigenvector with eigenvalue $\lambda^{*}$. For each pair of complex eigenvalues, the eigenoperator associated with the eigenvalue
$\lambda$ is given by
\begin{equation}
\widehat{\varPsi}_{n}\left(z\right)=\widehat{\psi}_{nx}(z)+\widehat{\psi}_{ny}^{\dagger}(z)
\label{eq:Ceigenv_mode1}
\end{equation}

\noindent where $\widehat{\psi}_{nx}(z)=\sum\limits_{m=1}^{N}x_{nm}\widehat{a}_{m}(z)$ and $\widehat{\psi}_{ny}(z)=y_{nm}\widehat{a}_{m}(z)$, and the eigenoperator associated with $\lambda^{*}$ is given by $\widehat{\varPsi}_{n}^{\dagger}\left(z\right)$. In this case, $\widehat{\varPsi}_{n}\left(z\right)$ is not a Hermitian operator. However, its linear combinations $\widehat{\varPsi}_{n}^{\dagger}\left(z\right)+\widehat{\varPsi}_{n}\left(z\right)$ and $i\left(\widehat{\varPsi}_{n}^{\dagger}\left(z\right)-\widehat{\varPsi}_{n}\left(z\right)\right)$ are Hermitian operators that could be characterized with the measurement setup depicted in Fig.~\ref{fig:General_theory}.

\section{Quantum gain and loss models in coupled waveguide systems}

Next, we use the described theoretical framework for the analysis of photonic systems composed of two coupled waveguides, one amplifying and one lossy waveguide. We aim to show how the linear or parametric nature of the gain and loss mechanisms influences the behavior of the eigenvalues, the evolution of the eigenoperators, and the photon
statistics of the noise generated by the quantum system.

\begin{figure*}[t]
\includegraphics[width=0.8\textwidth]{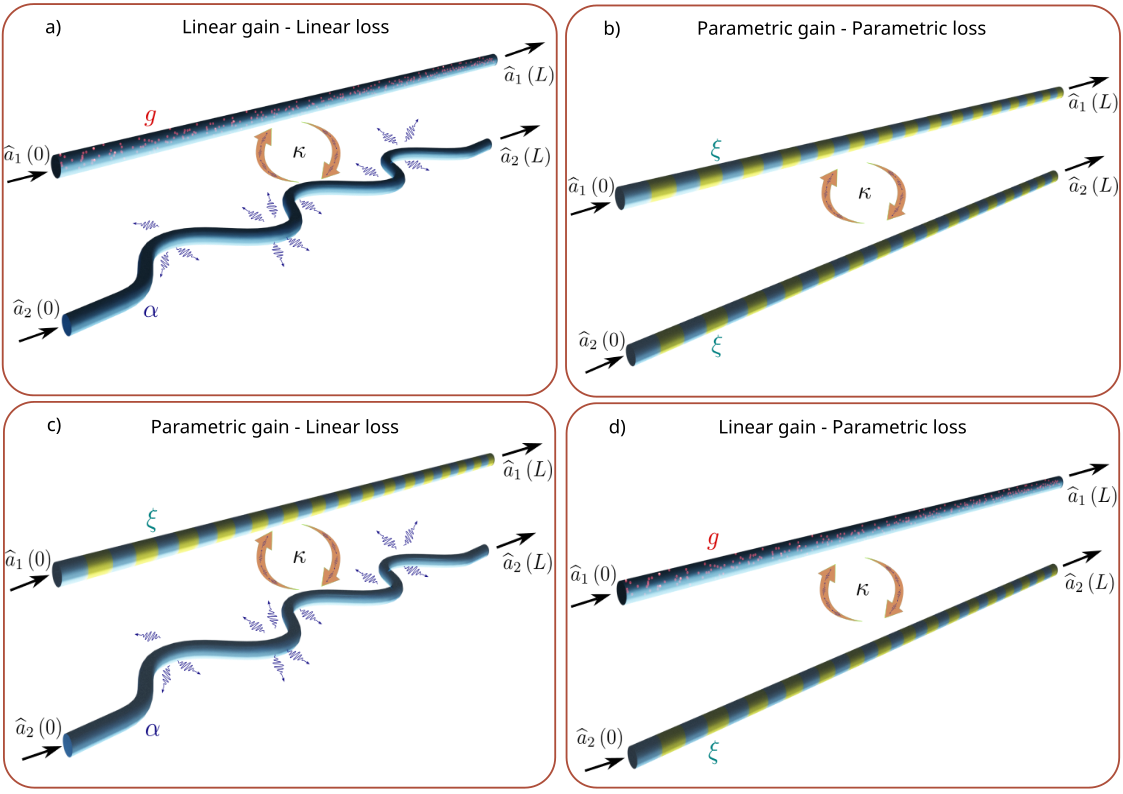}
\caption{\label{fig:Wvg_sketch}Sketch of coupled waveguides with amplification and losses modeled through a combination of linear and parametric phenomena.}
\end{figure*}

To this end, we consider four different scenarios, which are schematically depicted in Fig.~\ref{fig:Wvg_sketch}: both gain and losses modeled through linear processes, both modeled through a parametric process, and the two possible combinations of linear and parametric mechanisms. It is important to note that in the figure, linear losses are showcased as arising from waveguide bending for a comprehensible graphical representation. However, the analysis proposed in the manuscript is not specific or restricted to them. Additionally, parametric gain and losses are indistinguishable in the depicted waveguides, as the mechanism underlying both phenomena is the same. 


\subsection{Linear gain-Linear loss}

As depicted in Figure \ref{fig:Wvg_sketch}(a), we start by analyzing a linear amplifying waveguide with gain factor $g$, coupled with strength $\kappa$ to a linear lossy waveguide
with loss factor $\alpha$. The spatial evolution of the photonics modes in the system is described through the following coupled differential equations:
\begin{equation}
\partial_{z}\widehat{a}_{1}\left(z\right)=g\widehat{a}_{1}\left(z\right)+i\kappa\widehat{a}_{2}\left(z\right)+\sqrt{2g}\widehat{f}_{1}^{\dagger}\left(z\right)
\end{equation}
\begin{equation}
\partial_{z}\widehat{a}_{2}\left(z\right)=-\alpha\widehat{a}_{2}\left(z\right)+i\kappa\widehat{a}_{1}\left(z\right)+\sqrt{2\alpha}\widehat{f}_{2}\left(z\right)
\end{equation}

\noindent where the bosonic noise operators and their associated coefficients are responsible for maintaining the commutation relations. Equivalently, the system's spatial evolution can be compactly written as in Eq.~(\ref{eq:dif_evolution}) , from where we can straightforwardly identify the matrices $\mathbf{P}$ and $\mathbf{Q}$ in the block matrix $\mathbf{M}$:
\begin{equation}
\mathbf{P}=\left[\begin{array}{cc}
g & i\kappa\\
i\kappa & -\alpha
\end{array}\right]\quad\mathbf{Q}=\left[\begin{array}{cc}
0 & 0\\
0 & 0
\end{array}\right]
\end{equation}

As mentioned, linear phenomena do not involve mixing of creation and annihilation operators, and, therefore, $\mathbf{Q}=\boldsymbol{0}$. In addition, $\mathbf{M}^{T}=\mathbf{M}$, since the coupling between the waveguides is reciprocal. Next, computing eigenvalues and eigenvectors of this system implies to solve Eq.~(\ref{eq:rel1}) and Eq.~(\ref{eq:rel2}), which upon substitution of the symmetric matrix $\mathbf{P}$ and $\mathbf{Q}$ reduce to $\mathbf{P}\mathbf{x}=\lambda\,\mathbf{x}$ and $\mathbf{P}^{*}\mathbf{y}^{*}=\lambda\,\mathbf{y}^{*}$, respectively. From these relations, we can predict that if the matrix $\mathbf{P}$ has eigenvalue $\lambda$ with eigenvector $\mathbf{x}$, then the matrix $\mathbf{M}$ would share the same eigenvalue $\lambda$ with eigenvector $\begin{bmatrix}\mathbf{x} & \mathbf{0}\end{bmatrix}^{T}$, and also one eigenvalue $\lambda^{*}$ with eigenvector $\begin{bmatrix}\mathbf{0} & \mathbf{x}^{*}\end{bmatrix}^{T}$, confirming that complex eigenvalues appear on complex conjugate pairs. At the same time, we find that real eigenvalues appear with duplicity 2. Therefore, the $N$-dimensional eigenvectors of the matrix $\mathbf{P}$ will define the $2N$-dimensional eigenvectors of the matrix $\mathbf{M}$. For the sake of simplicity and to compare to the classical analogue of two coupled waveguides with balanced gain and loss, we consider the particular case where $g=\alpha$. Therefore, we obtain two pairs of degenerated eigenvalues (four eigenvalues):
\begin{equation}
\lambda_{1,2\,\pm}=\pm\sqrt{\alpha^{2}-\kappa^{2}}\label{eq:Eigenval_LinGL}
\end{equation}

In the notation adopted, the $\pm$ sign refers to the positive or negative square root, while the sub-index $1$ ($2$) refers to eigenvalues or eigenvectors resulting from solutions to Eq.~(\ref{eq:rel1}) (Eq.~(\ref{eq:rel2})) in the reduced form described above. Therefore, we have eigenvalues $\lambda_{1\pm}$ from Eq.~(\ref{eq:rel1}) and $\lambda_{2\pm}$ from Eq.~(\ref{eq:rel2}). The associated normalized eigenvectors are given by
\begin{equation}
\mathbf{v}_{1\pm}=\frac{1}{\sqrt{1+\left|\beta_{\pm}\right|^{2}}}\begin{bmatrix}1 & i\beta_{\pm} & 0 & 0\end{bmatrix}^{T}
\label{eq:LinGL_generalEigenvect1}
\end{equation}
\begin{equation}
\mathbf{v}_{2\pm}=\frac{1}{\sqrt{1+\left|\beta_{\pm}\right|^{2}}}\begin{bmatrix}0 & 0 & 1 & -i\beta_{\pm}\end{bmatrix}^{T}
\label{eq:LinGL_generalEigenvect2}
\end{equation}

\noindent where $\beta_{\pm}=-\frac{1}{\kappa}\left(-\alpha\pm\sqrt{\alpha^{2}-\kappa^{2}}\right)$.


\begin{figure*}[t]
\includegraphics[width=0.8
\textwidth]{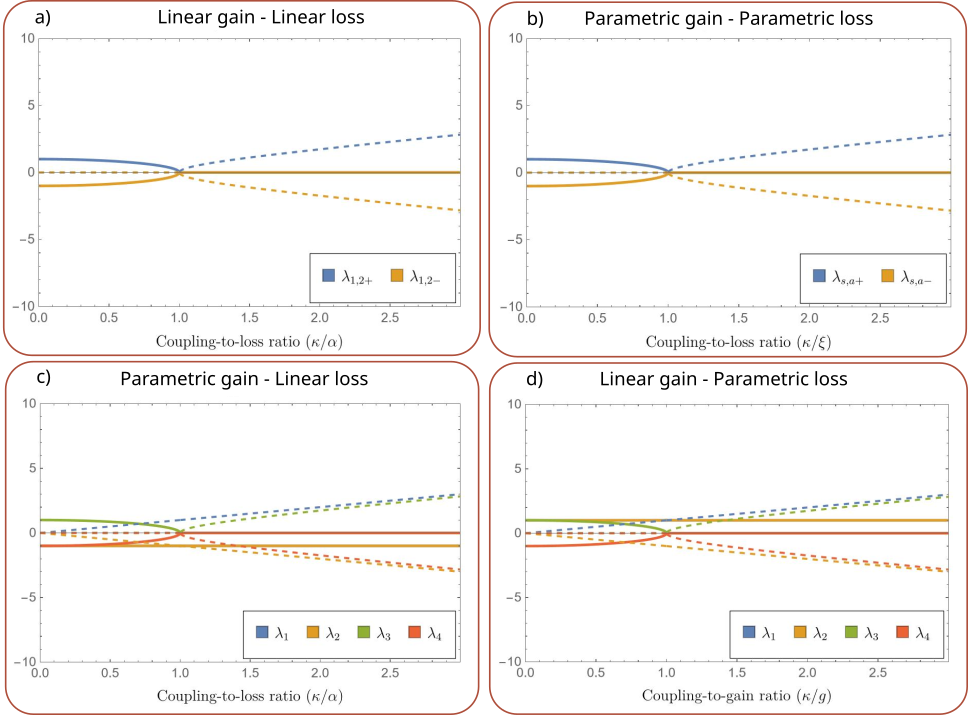}
\caption{\label{fig:Eigenvalues}Real (solid lines) and imaginary parts (dashed lines) of the eigenvalues in coupled waveguides with balanced gain and loss modeled through a combination of linear and parametric phenomena.}
\end{figure*}

As noticed from (\ref{eq:Eigenval_LinGL}), the eigenvalues will be real or complex numbers depending on the interplay between the coupling coefficient $\kappa$ and the loss (gain) coefficient $\alpha$. It is clear from Figure \ref{fig:Eigenvalues}(a) that there is a phase transition point in the parameter space, determined by the coupling to loss ratio $\nicefrac{\kappa}{\alpha}=1$, where all the eigenvalues coalesce and vanish, i.e., $\lambda_{1,2\,\pm}=0$, establishing the transition from a phase with real eigenvalues to another phase where all the eigenvalues are complex numbers. At this transition point $\beta_{\pm}=1$ and the space expanded by the eigenvectors shrinks because they coalesce in pairs, with $\mathbf{v}_{1\pm}=\frac{1}{\sqrt{2}}\begin{bmatrix}1 & i & 0 & 0\end{bmatrix}^{T}$ and $\mathbf{v}_{2\pm}=\frac{1}{\sqrt{2}}\begin{bmatrix}0 & 0 & 1 & -i\end{bmatrix}^{T}$.  At the eigenvalue/eigenvector level, the linear gain / linear loss configuration has the clearest connection with the classical description of the system. From a classical perspective, this phase transition point corresponds to an exceptional point (EP), where all eigenvalues and eigenvectors coalesce \cite{wang2023non}. From the quantum optics
perspective, the distinction between creation and destruction operators, doubles the eigenspace dimension, and not all eigenvectors coalesce due to complex conjugation. In addition, the inclusion of noise terms also distinguishes the quantum approach \cite{scheel2018symmetric}.

\subsubsection{Eigenvectors and eigenoperators for real eigenvalues}

Our analysis focuses on the case of real eigenvalues of $\mathbf{M}$ leading to amplification and attenuation effects, and how they converge to the phase transition, also leading to gain-loss compensation. As anticipated by Eq.~(\ref{eq:Eigenval_LinGL}) and Figure \ref{fig:Eigenvalues}(a), if $\alpha>\kappa$ we have that all eigenvalues are real and the
eigenvectors can be compactly written as in (\ref{eq:LinGL_generalEigenvect1}) and (\ref{eq:LinGL_generalEigenvect2}) ignoring the module operation since $\beta_{\pm}$ would only take real values. We can verify that in the limit $\kappa\rightarrow0$, the eigenvalues become $\lambda_{1,2\pm}=\pm\sqrt{\alpha^{2}-\kappa^{2}}\simeq\pm\alpha$. Additionally, $\beta_{+}=0$ and $\beta_{-}\rightarrow\infty$, and therefore, we recover the limit case of two uncoupled amplifying and lossy waveguides.

As previously explained, eigenvectors associated with real eigenvalues can be represented in a way that leads to Hermitian eigenoperators describing the system's modes. As the $+$ and $-$ eigenvectors have the same eigenvalue with multiplicity two, it is possible to write a different basis that complies with the required structure $\boldsymbol{v}=\begin{bmatrix}\mathbf{x} & \mathbf{x}^{*}\end{bmatrix}^{T}$ by defining two quadratures:
\begin{equation}
\mathbf{v}_{X\pm}=\frac{1}{2}\left(\mathbf{v}_{1\pm}+\mathbf{v}_{2\pm}\right)=\frac{1}{2\sqrt{1+\beta_{\pm}^{2}}}\begin{bmatrix}1 & i\beta_{\pm} & 1 & -i\beta_{\pm}\end{bmatrix}^{T}
\end{equation}
\begin{equation}
\mathbf{v}_{Y\pm}=\frac{i}{2}\left(\mathbf{v}_{2\pm}-\mathbf{v}_{1\pm}\right)=\frac{1}{2\sqrt{1+\beta_{\pm}^{2}}}\begin{bmatrix}-1 & -i\beta_{\pm} & 1 & -i\beta_{\pm}\end{bmatrix}^{T}
\end{equation}

Therefore, the associated Hermitian eigenoperators can be defined as follows:
\begin{multline}
\widehat{\varPsi}_{X\pm}(L)=\frac{1}{2}\left(\widehat{a}_{\pm}(0)+\widehat{a}_{\pm}^{\dagger}(0)\right)e^{\lambda_{\pm}L}\\
\qquad+\sqrt{2\alpha}\,e^{\lambda_{\pm}L}\int_{0}^{L}dz'\,\frac{1}{2}\left(\widehat{f}_{\pm}(z')+\widehat{f}_{\pm}^{\dagger}(z')\right)e^{-\lambda_{\pm}z'}
\end{multline}
\begin{multline}
\widehat{\varPsi}_{Y\pm}(L)=\frac{i}{2}\left(\widehat{a}_{\pm}^{\dagger}(0)-\widehat{a}_{\pm}(0)\right)e^{\lambda_{\pm}L}\\
\qquad+\sqrt{2\alpha}\,e^{\lambda_{\pm}L}\int_{0}^{L}dz'\,\frac{i}{2}\left(\widehat{f}_{\pm}^{\dagger}(z')-\widehat{f}_{\pm}(z')\right)e^{-\lambda_{\pm}z'}
\end{multline}

\noindent where $\widehat{a}_{\pm}(0)=\frac{\widehat{a}_{1}(0)+i\beta_{\pm}\widehat{a}_{2}(0)}{\sqrt{1+\beta_{\pm}^{2}}}$ and $\widehat{f}_{\pm}(z')=\frac{\widehat{f}_{1}(z')+i\beta_{\pm}\widehat{f}_{2}(z')}{\sqrt{1+\beta_{\pm}^{2}}}$.

\subsubsection{Eigenoperators variance}

Next we study the noise properties in the system by computing the variance for each quadrature eigenoperator:
\noindent 
\begin{equation}
\left(\Delta\varPsi_{n}(L)\right)^{2}=\left\langle \widehat{\varPsi}_{n}^{2}(L)\right\rangle -\left\langle \widehat{\varPsi}_{n}(L)\right\rangle ^{2}\label{eq:variance}
\end{equation}

\noindent where the angle brackets $\left\langle \right\rangle $ denote expectations values referred to the quantum state of the system, at the beginning of the waveguide. We are interested in the noise inherent to the device, due to the nature of the gain and loss mechanisms and the thermal conditions in the system. Therefore, in all the systems analyzed in this manuscript, the expectation values computed refer to the joint state describing input vacuum states in both waveguides and also the waveguides at a given temperature $T$.
Such conditions can be described through the joint density matrix $\hat{\rho}=\hat{\rho_{1}}\otimes\hat{\rho_{2}}$ with $\hat{\rho}_{1,2}=\left(1-e^{-\nicefrac{\hbar\omega}{k_{B}T_{1,2}}}\right)\underset{n_{f1}}{\sum}e^{-\nicefrac{n_{f1,2}\hbar\omega}{k_{B}T_{1,2}}}\left|n_{f1,2}\right\rangle \left\langle n_{f1,2}\right|$, where $\hbar$ is the reduced Planck constant, $k_{B}$ is the Boltzmann constant and $n_{fn}$ represents the number of thermal photons in mode $n$. Here, it is important to remark that if the linear loss of the waveguide correspond to dissipation loss, then the temperature $T$ physically corresponds to the temperature of the system. However, for scattering and/or radiation loss, then
the temperature $T$ physically corresponds to an effective temperature describing the external background noise coupled to the waveguide.

As the eigenoperators involved only contain linear combinations of annihilation and creation operators, the expectation values of the individual operators vanish, and the contribution to the variance is only due to the squared operator expectation value, i.e., $\left(\Delta\varPsi_{n}(L)\right)^{2}=\left\langle \widehat{\varPsi}_{n}^{2}(L)\right\rangle $. Additionally, it is easy to verify that the $+$ and $-$ eigenoperators in both quadratures present the same statistics $\left(\Delta\varPsi_{Y\pm}(L)\right)^{2}=\left(\Delta\varPsi_{X\pm}(L)\right)^{2}$, which is reasonable considering the absence of phase-sensitive gain or loss mechanisms that influence differently each quadrature in the
system, as it is the case in a parametric process represented through a squeezing transformation. Therefore, we have 
\begin{multline}
\left(\Delta\varPsi_{X,Y\pm}(L)\right)^{2}=\frac{1}{4}e^{2\lambda_{\pm}L}\\
\qquad+\frac{\alpha\,\left(e^{2\lambda_{\pm}L}-1\right)}{4\left(1+\beta_{\pm}^{2}\right)\lambda_{\pm}}\left(2\left\langle \widehat{n}_{f1}\right\rangle +2\beta_{\pm}^{2}\left\langle \widehat{n}_{f2}\right\rangle +1+\beta_{\pm}^{2}\right)
\label{eq:LinGL_var}
\end{multline}

\noindent where $\widehat{n}_{fn}=\widehat{f}_{n}^{\dagger}(0)\,\widehat{f}_{n}(0)$ represents the number operator for thermal photons in mode $n$. The mean number of photons in a thermal state \cite{loudon2000quantum} is given by $\left\langle \hat{n}\right\rangle _{T}=\frac{1}{e^{\nicefrac{\hbar\omega}{k_{B}T}}-1}$. It is easy to distinguish that the first term in Eq.~(\ref{eq:LinGL_var}) corresponds to the contribution from the photonic part in the eigenoperator, while the second term is contributed by the noise.


\begin{figure*}[t]
\includegraphics[width=0.8\textwidth]{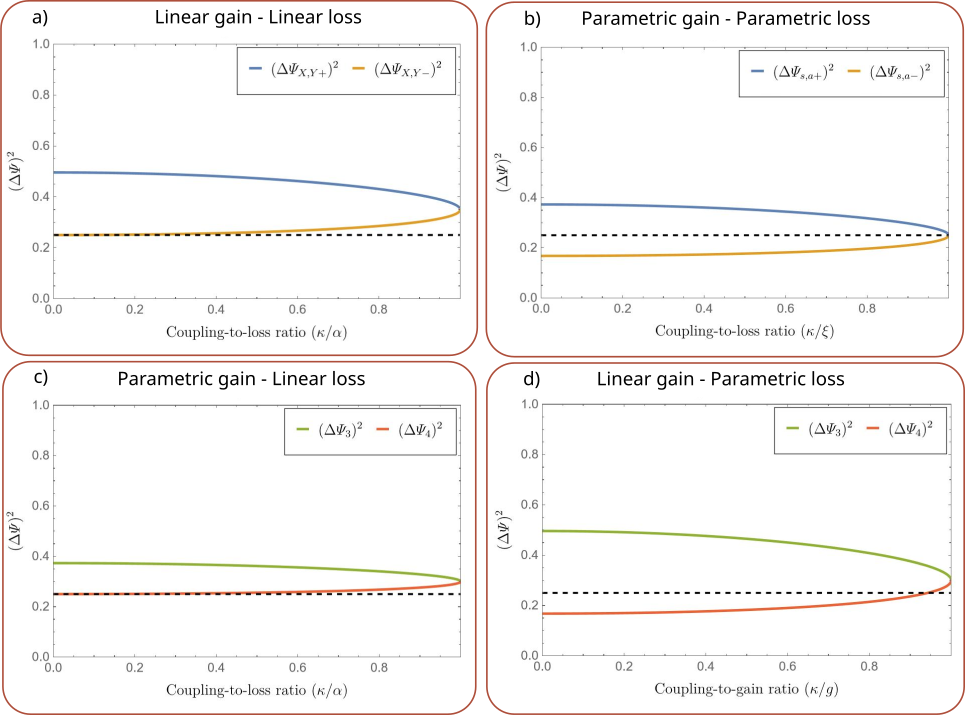}
\caption{\label{fig:Variance_vsCoup2Loss}Eigenoperators variance as a function of the coupling-to-loss (or gain) ratio, considering input vacuum states and waveguides at temperature $T=0$. The plots correspond to the case when $\ensuremath{\alpha\,L}=0.2$.}
\end{figure*}

It is interesting to analyze some limit cases. For instance, in the zero temperature limit, $T=0$, the variance reduces to
\begin{equation}
\left(\Delta\varPsi_{X,Y\pm}(L)\right)_{T0}^{2}=\frac{1}{4}e^{2\lambda_{\pm}L}+\frac{\alpha}{4\lambda_{\pm}}\left(e^{2\lambda_{\pm}L}-1\right)\label{eq:LinGL_var_T0}
\end{equation}

These results are plotted in Figure \ref{fig:Variance_vsCoup2Loss}(a), as a function of the coupling to loss ratio ($\kappa/\alpha$) for a given $\text{\ensuremath{\alpha\,L}}$. As expected, in the limit $\kappa\rightarrow0$, where $\lambda_{\pm}\rightarrow\pm\alpha$, we recover the variances associated with uncoupled waveguides, i.e., $\left(\Delta\varPsi_{X,Y+}(L)\right)^{2}=\frac{1}{4}\left\{ 2e^{2\alpha L}-1\right\} $ for a waveguide with linear gain and $\left(\Delta\varPsi_{X,Y-}(L)\right)^{2}=\frac{1}{4}$ for the lossy waveguide. When we move away from this limit, a stronger coupling between photons in different waveguide modes causes the variance of the amplified modes to decrease, while the lossy modes increase their variance, until they balance $\left(\Delta\varPsi_{X,Y\pm}(L)\right)_{\kappa=\alpha}^{2}=\frac{1}{4}+\frac{1}{2}\alpha L$ at the phase transition point where the eigenvalues vanish. At the phase transition, linear gain and loss are perfectly balanced and classical signals are neither amplified nor attenuated. However, it is found that the system introduces additional noise that increases the variance in both quadratures. The larger the loss compensation,
the stronger the noise. Remarkably, the generated noise at the phase transition only scales linearly with the length of the waveguide, a much slower trend that the exponential scaling of uncoupled ($\kappa\rightarrow0$) waveguides. In fact, it can be demonstrated that the variance at the phase transition corresponds to the geometric mean of the variances of the uncoupled waveguides for sufficiently small values of $\alpha L$. In other words, even if the waveguides are electrically large, the variance of the generated
noise at the phase transition corresponds to taking the Taylor series of the geometric mean of the variances of the noise in the uncoupled waveguides, and keeping only the first order approximation. Therefore, although noise is unavoidably generated at the gain-loss compensation phase transition point, its scaling is much slower than that exhibited
by uncoupled ($\kappa\rightarrow0$) lossy and gain waveguides. 

\begin{figure*}[t]
\includegraphics[width=0.8\textwidth]{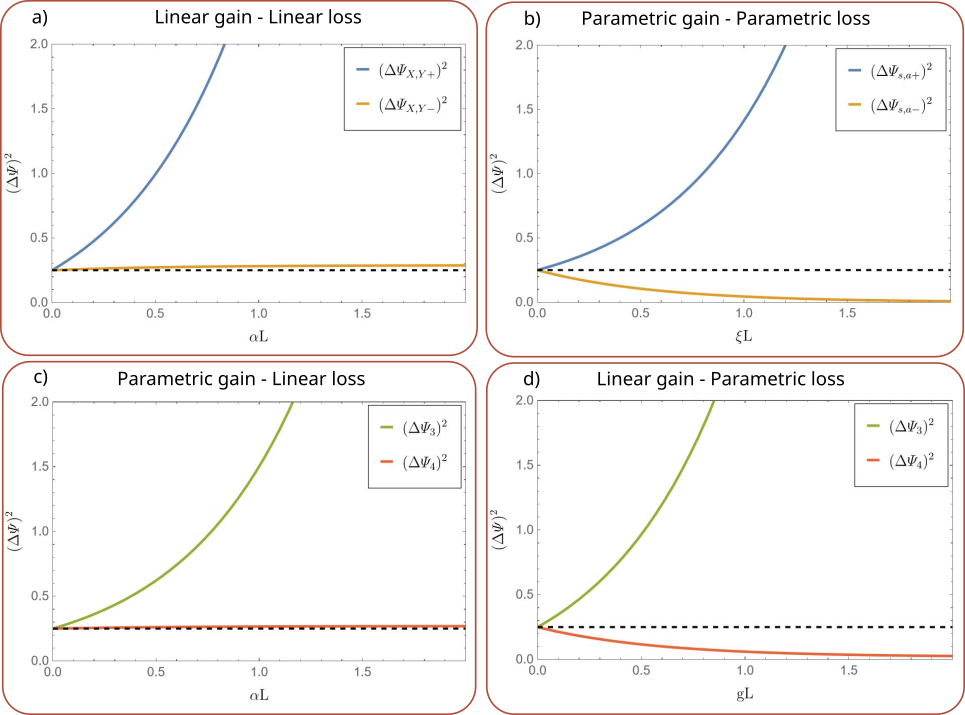}
\caption{\label{fig:Variance_vsL}Dependence of the eigenoperators variance on the waveguide length, considering input vacuum states and waveguides at temperature $T=0$. The coupling-to-loss (or gain) ratio has been set equal to $0.5$}
\end{figure*}

Moreover, plotting (\ref{eq:LinGL_var_T0}) as a function of the waveguide length $L$ for a given coupling to loss ratio within the limits for real eigenvalues ($\kappa/\alpha=0.5$) as in Figure \ref{fig:Variance_vsL}(a), confirms that increasing $L$ leads to higher variance values. However, as we can deduce from Eq.~(\ref{eq:LinGL_var_T0}), eigenmodes associated with different eigenvalues are influenced differently; therefore, one pair experiences an evident exponential growth in the fluctuations, while the other pair is almost unaffected.

\begin{figure*}[t]
\includegraphics[width=0.8\textwidth]{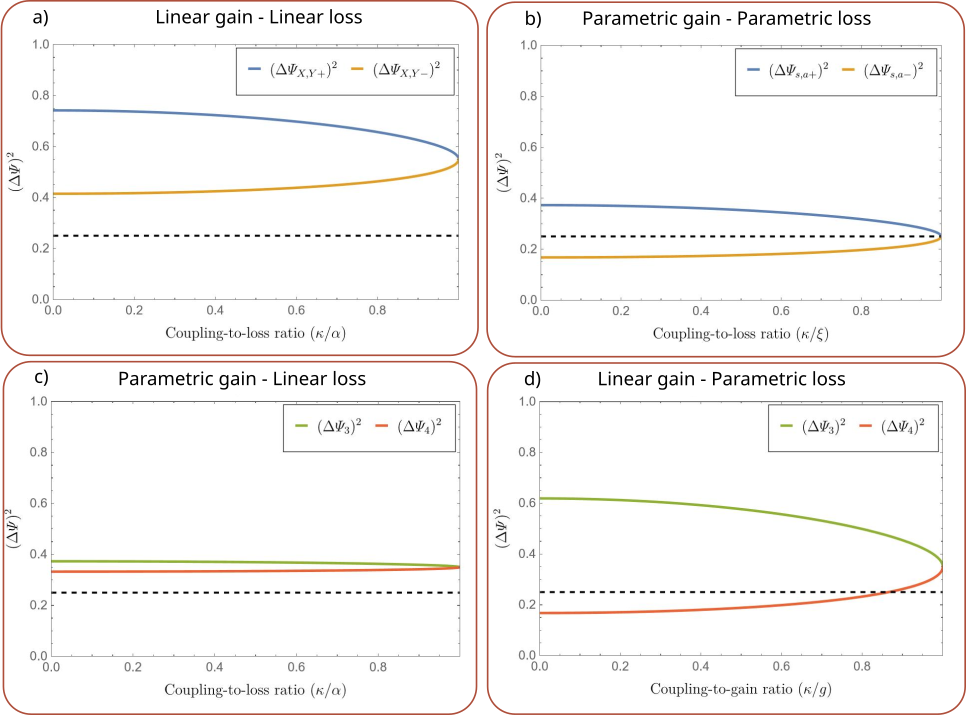}
\caption{\label{fig:Variance_vsCoup2Loss_1nf}Eigenoperators variance as a function of the coupling-to-loss (or gain) ratio, considering input vacuum states and one thermal photon in each waveguide. The plots correspond to the case when $\ensuremath{\alpha\,L}=0.2$.}
\end{figure*}

On the other hand, assuming that both waveguides are at the same temperature $T\neq0$, the variance is given by
\begin{multline}
\left(\Delta\varPsi_{X,Y\pm}(L)\right)_{T}^{2}=\frac{1}{4}e^{2\lambda_{\pm}L}\\
\qquad+\frac{\alpha}{4\lambda_{\pm}}\left(e^{2\lambda_{\pm}L}-1\right)\left(2\left\langle \widehat{n}_{f}\right\rangle +1\right)
\end{multline}

\noindent where we can straightforwardly identify the thermal noise contribution, as $\left(\Delta\varPsi_{X,Y\pm}(L)\right)_{T}^{2}=\left(\Delta\varPsi_{X,Y\pm}(L)\right)_{T0}^{2}+\frac{\alpha\left(e^{2\lambda_{\pm}L}-1\right)}{2\lambda_{\pm}}\left\langle \widehat{n}_{f}\right\rangle $. Figure \ref{fig:Variance_vsCoup2Loss_1nf} (a) depicts the new results
for one thermal photon $\left\langle \widehat{n}_{f}\right\rangle =1$ in each waveguide. We observe a significant increase of the fluctuations in both modes even for a single thermal photon.


It is important to recall that if $\alpha=\kappa$ we obtain that all the four eigenvalues vanish $\lambda_{1,2\pm}=0$. Additionally, $\beta_{\pm}=1$ and there is a coalescence in pairs of the eigenvectors with $\mathbf{v}_{X\pm}=\frac{1}{2\sqrt{2}}\begin{bmatrix}1 & i & 1 & -i\end{bmatrix}^{T}$ and $\mathbf{v}_{Y\pm}=\frac{1}{2\sqrt{2}}\begin{bmatrix}-1 & -i & 1 & -i\end{bmatrix}^{T}$. Computing the variance on the quadrature basis leads to $\left(\Delta\varPsi_{X,Y\pm}(L)\right)_{\kappa=\alpha}^{2}=\frac{1}{4}+\frac{1}{2}\alpha L\left(\left\langle \widehat{n}_{f1}\right\rangle +\left\langle \widehat{n}_{f2}\right\rangle +1\right)$, which is the same result we recover from Eq.~(\ref{eq:LinGL_var})
in the limit $\lambda_{\pm}\rightarrow0$, pointing towards continuity at the phase transition point. Therefore, both thermal and quantum noise scale linearly with the length of the waveguide at the phase transition point.

\subsection{Parametric gain-Parametric loss}

The next waveguide system under analysis consist also in one amplifying and one lossy waveguides coupled with strength $\kappa$, as represented in Figure \ref{fig:Wvg_sketch} (b). Gain and losses are modeled through the same parametric process: a squeezing transformation with squeezing parameter $\xi\in\mathbb{R}$, which is considered a real quantity for simplicity. For this configuration, the spatial evolution of the photonics modes in the system is given by \cite{loudon2000quantum}:
\begin{equation}
\partial_{z}\widehat{a}_{1}\left(z\right)=\xi\widehat{a}_{1}^{\dagger}\left(z\right)+i\kappa\widehat{a}_{2}\left(z\right)
\end{equation}
\begin{equation}
\partial_{z}\widehat{a}_{2}\left(z\right)=\xi\widehat{a}_{2}^{\dagger}\left(z\right)+i\kappa\widehat{a}_{1}\left(z\right)
\end{equation}

\noindent Again, the system can be equivalently described as in Eq.~(\ref{eq:dif_evolution}), with $\mathbf{P}$ and $\mathbf{Q}$ matrices:
\begin{equation}
\mathbf{P}=\left[\begin{array}{cc}
0 & i\kappa\\
i\kappa & 0
\end{array}\right],\,\,\,\,\,\mathbf{Q}=\left[\begin{array}{cc}
\xi & 0\\
0 & \xi
\end{array}\right]
\end{equation}

Using a common mechanism for both gain and loss makes it possible to interchange the role of the amplifying and lossy waveguides, endowing the system with an inherent symmetry. Therefore, a natural basis to analyze the system is the use of symmetric, $\boldsymbol{v}_{s}=\begin{bmatrix}x & x & y^{*} & y^{*}\end{bmatrix}^{T}$, and anti-symmetric, $\boldsymbol{v}_{a}=\begin{bmatrix}x & -x & y^{*} & -y^{*}\end{bmatrix}^{T}$, eigenvectors. Then, substituting these constraints into Eq.~(\ref{eq:rel1}) and Eq.~(\ref{eq:rel2}) we find that both symmetric and anti-symmetric eigenvectors share the same pair of eigenvalues
\begin{equation}
\lambda_{s,a\,\pm}=\pm\sqrt{\xi^{2}-\kappa^{2}}
\label{eq:NonGL_eigenval}
\end{equation}

\noindent which can be real or complex numbers depending on the interplay between squeezing and coupling parameters. A phase transition occurs at $\nicefrac{\kappa}{\xi}=1$, where all eigenvalues simultaneously coalesce and vanish $\lambda_{s,a\,\pm}=0$, as depicted in Figure \ref{fig:Eigenvalues} (b). It is clear from Eq.~(\ref{eq:NonGL_eigenval}) and Figure \ref{fig:Eigenvalues} (b) that the eigenvalues behave exactly the same as in the linear gain - linear loss case, with the squeezing parameter $\xi$ playing the role of the loss (gain) coefficient $\alpha$ in (\ref{eq:Eigenval_LinGL}). At the same time, the eigenvectors and the generated noise have very different properties.

\subsubsection{Eigenvectors and eigenoperators for real eigenvalues}

Hereafter, we will focus on the regime where the degenerate eigenvalues are real numbers, i.e. $\xi\geq\kappa$. For real eigenvalues, the eigenvectors take the form
\begin{equation}
v_{s\pm}=\frac{1}{2}e^{i\nicefrac{\varphi_{s\pm}}{2}}\begin{bmatrix}e^{-i\nicefrac{\varphi_{s\pm}}{2}} & e^{-i\nicefrac{\varphi_{s\pm}}{2}} & e^{i\nicefrac{\varphi_{s\pm}}{2}} & e^{i\nicefrac{\varphi_{s}\pm}{2}}\end{bmatrix}^{T}
\end{equation}
\begin{equation}
v_{a\pm}=\frac{1}{2}e^{i\nicefrac{\varphi_{a\pm}}{2}}\begin{bmatrix}e^{-i\nicefrac{\varphi_{a\pm}}{2}} & -e^{-i\nicefrac{\varphi_{a\pm}}{2}} & e^{i\nicefrac{\varphi_{a\pm}}{2}} & -e^{i\nicefrac{\varphi_{a\pm}}{2}}\end{bmatrix}^{T}
\end{equation}

\noindent where $e^{i\varphi_{s\pm}}=\frac{\pm\sqrt{\xi^{2}-\kappa^{2}}-i\kappa}{\xi}$ and $e^{i\varphi_{a\pm}}=\frac{\pm\sqrt{\xi^{2}-\kappa^{2}}+i\kappa}{\xi}$, and therefore, $\varphi_{a\pm}=\arctan\frac{\kappa}{\pm\sqrt{\xi^{2}-\kappa^{2}}}$ and $\varphi_{s\pm}=\arctan\frac{-\kappa}{\pm\sqrt{\xi^{2}-\kappa^{2}}}$. The $\pm$ sign in the phase angle represents the positive or negative square root in the eigenvalue computation. The extracted common phase-factors ensures that excluding this global phase, the eigenvectors take the predicted form for real eigenvalues, leading to Hermitian operators. At the eigenvalues phase-transition point $\nicefrac{\kappa}{\xi}=1$, we have that $e^{i\varphi_{s\pm}}=-i$ and $e^{i\varphi_{a\pm}}=i$, and the eigenvectors associated to the $+$ and $-$ eigenvalues in each mode coalesce. Therefore, at the phase transition point where the four eigenvalues vanish $\lambda_{s,a\,\pm}=0$, the eigenvectors only coalesce in pairs. In this case, the phase sensitivity of the system precludes the existence of a EP where all eigenvalues and eigenvectors coalesce.

Finally, the eigenoperators at a distance $L$ on the waveguides are given by
\begin{equation}
\widehat{\varPsi}_{s\pm}(L)=\frac{1}{2}e^{i\nicefrac{\varphi_{s\pm}}{2}}\left[\widehat{a}_{s}(0)+\widehat{a}_{s}^{\dagger}(0)\right]e^{\lambda_{s,a\,\pm}L}
\end{equation}
\begin{equation}
\widehat{\varPsi}_{a\pm}(L)=\frac{1}{2}e^{i\nicefrac{\varphi_{a\pm}}{2}}\left[\widehat{a}_{a}(0)+\widehat{a}_{a}^{\dagger}(0)\right]e^{\lambda_{s,a\,\pm}L}
\end{equation}

\noindent where $\widehat{a}_{s}(0)=e^{-i\nicefrac{\varphi_{s\pm}}{2}}\,\frac{\widehat{a}_{1}(0)+\widehat{a}_{2}(0)}{\sqrt{2}}$ and $\widehat{a}_{a}=e^{-i\nicefrac{\varphi_{a\pm}}{2}}\,\frac{\widehat{a}_{1}(0)-\widehat{a}_{2}(0)}{\sqrt{2}}$.

\subsubsection{Eigenoperators variance}

Computing the variance of the eigenoperators considering vacuum states in both waveguides and both at the same temperature $T$ reveals important aspects of the fluctuations in the coupled system:
\begin{equation}
\left(\Delta\varPsi_{s,a\pm}(L)\right)_{T}^{2}=\frac{1}{4}e^{2\lambda_{s,a\,\pm}L}
\label{eq:NonGL_var}
\end{equation}

First, as expected from the absence of noise sources in the evolution equations, there is no thermal noise contribution to the system variance; therefore, the system is only affected by the unavoidable quantum noise that results from vacuum fluctuations. Second, for both the symmetric and anti-symmetric modes, the variance behaves similarly
to the variance in a squeezing transformation. The difference is that instead of twice the squeezing amplitude in the exponential ($\pm2\xi$) we now have this term replaced by $2\lambda_{\pm}L=\text{\ensuremath{\pm}}2\sqrt{\xi^{2}-\kappa^{2}}\,L$, which can be interpreted as a modified squeezing amplitude. It suggests that increasing the coupling strength between the waveguides decreases the degree of squeezing we can obtain, while a larger interaction length would have the opposite effect. At the uncoupled waveguides
limit, i.e. $\kappa\to0$, the variance reduces to $\left(\Delta\varPsi_{s,a\pm}(L)\right)^{2}=\frac{1}{4}e^{\pm2\xi L}$. Figure \ref{fig:Variance_vsCoup2Loss} confirms that a pair of eigenmodes exhibit a squeezed variance with a minimum value when $\kappa\to0$, which increases with the coupling-to-loss ratio $\nicefrac{\kappa}{\xi}$ but remains squeezed until it reaches the vacuum fluctuation limit $\nicefrac{1}{4}$ at the critical phase-transition point. As for the anti-squeezed modes, their variance reduces from its maximum value for the uncoupled waveguides limit until it also converges to the vacuum fluctuation limit $\nicefrac{1}{4}$ at the phase-transition point. As explained, the evolution of the eigenmodes and their variance are not affected by the temperature of the device, as shown in Figure \ref{fig:Variance_vsCoup2Loss_1nf} (b), that coincides with Figure \ref{fig:Variance_vsCoup2Loss}(b). At the transition point gain and loss are perfectly compensated, and their squeezing is nil. At the same time, such gain loss compensation occurs without any noise amplification. Thus, the variance experienced by the photonic modes at the phase transition point correspond to unamplified vacuum fluctuations.

Regarding the dependence of the eigenoperator's variance on the waveguide length $L$, as anticipated from Eq. (\ref{eq:NonGL_var}) and depicted in Figure \ref{fig:Variance_vsL} (b), we observe an exponential reduction of the fluctuations in the squeezed modes and a faster exponential increase in the anti-squeezed modes, both starting from the vacuum value at the infinitesimal waveguide limit.

\subsection{Parametric gain-Linear loss}

After analyzing systems of two coupled waveguides exhibiting both linear gain and loss or both parametric gain and loss, a natural question arises about the effect of combining linear and nonlinear phenomena. Therefore, we study a system composed by an amplifying waveguide, with parametric gain modeled through a squeezing transformation with squeezing parameter $\xi$, coupled with strength $\kappa$ to a linear lossy waveguide with loss factor $\alpha$, as depicted in Figure \ref{fig:Wvg_sketch} (c). The spatial evolution of the photonic modes in the system can be written as follows:
\begin{equation}
\partial_{z}\widehat{a}_{1}\left(z\right)=\xi\widehat{a}_{1}^{\dagger}\left(z\right)+i\kappa\widehat{a}_{2}\left(z\right)
\end{equation}
\begin{equation}
\partial_{z}\widehat{a}_{2}\left(z\right)=-\alpha\widehat{a}_{2}\left(z\right)+i\kappa\widehat{a}_{1}\left(z\right)+\sqrt{2\alpha}\widehat{f}_{2}\left(z\right)
\end{equation}

\noindent or, equivalently
\begin{equation}
\mathbf{P}=\left[\begin{array}{cc}
0 & i\kappa\\
i\kappa & -\alpha
\end{array}\right]\quad\mathbf{Q}=\left[\begin{array}{cc}
\xi & 0\\
0 & 0
\end{array}\right]
\end{equation}

\noindent and upon substitution into Eq.~(\ref{eq:rel1}) and Eq.~(\ref{eq:rel2}), we compute the eigenvalues for the system with balanced gain and loss $\alpha=\xi$:
\begin{equation}
\lambda_{1,2}=-\alpha\pm i\kappa
\label{eq:loss-loss_eigenval}
\end{equation}
\begin{equation}
\lambda_{3,4}=\pm\sqrt{\alpha^{2}-\kappa^{2}}
\label{eq:NonGLinL_eigenval}
\end{equation}

From the computed eigenvalues and their dependence on the coupling-to-loss ratio depicted in Figure \ref{fig:Eigenvalues} (c), we note that real-valued $\lambda_{1,2}$ will not occur for the balanced system under analysis; otherwise, there should be no coupling between waveguides. In fact, the eigenvalues in (\ref{eq:loss-loss_eigenval}) coincide
with those in a linear loss-loss configuration. The other pair of eigenvalues $\lambda_{3,4}$ coincides with the eigenvalues for the cases of linear gain - linear loss and parametric gain - parametric loss previously addressed. Therefore, $\lambda_{3,4}$ might take real or complex values depending on the interplay between loss (or gain) and coupling strength, with $\lambda_{3,4}\in\mathbb{C}$ for $\alpha<\kappa$, while $\lambda_{3,4}\in\mathbb{R}$ for $\alpha\geq\kappa$, being $\alpha=\kappa$ the phase-transition point ($\lambda_{3,4}=0$). Thus, we find that different classes of gain and loss can compensate each other, and they follow the same eigenvalue structure.

\subsubsection{Eigenvectors and eigenoperators for real eigenvalues }

As we are interested in gain-loss configurations, we focus on the pair of eigenvalues $\lambda_{3,4}$. The associated normalized eigenvectors for $\alpha>\kappa$ can be compactly written as follows:
\begin{equation}
\boldsymbol{v}_{3,4}=\frac{1}{\sqrt{1+A_{3,4}^{2}}}\begin{bmatrix}-iA_{3,4}, & 1, & iA_{3,4}, & 1\end{bmatrix}^{T}
\label{eq:NonlGLinL_eigenvect}
\end{equation}

\noindent with $A_{3,4}=\frac{\alpha\pm\sqrt{\alpha^{2}-\kappa^{2}}}{k}$. At the eigenvalues' phase-transition point we have that $A_{3}=A_{4}$ and then both eigenvectors coalesce. The associated eigenoperators are
\begin{multline}
\widehat{\varPsi}_{3,4}\left(L\right)=\frac{1}{2}\left(\widehat{a}_{3,4}(0)+\widehat{a}_{3,4}^{\dagger}(0)\right)e^{\lambda_{3,4}L}+\\
\frac{\sqrt{2\alpha}}{\sqrt{1+A_{3,4}^{2}}}e^{\lambda_{3,4}L}\int_{0}^{L}dz'e^{-\lambda_{3,4}z'}\frac{1}{2}\left(\widehat{f}_{2}(z')+\widehat{f}_{2}^{\dagger}(z')\right)
\end{multline}

\noindent where $\widehat{a}_{3,4}(0)=\frac{-iA_{3,4}\widehat{a}_{1}(0)+\widehat{a}_{2}(0)}{\sqrt{1+A_{3,4}^{2}}}$.

\subsubsection{Eigenoperators variance}

The variance characterizing the modes in the output of the waveguides can be computed as follows
\begin{multline}
\left(\Delta\varPsi_{3,4}(L)\right)_{T}^{2}=\frac{1}{4}e^{2\lambda_{3,4}L}+\\
\frac{1}{\left(1+A_{3,4}^{2}\right)}\frac{\alpha}{4\lambda_{3,4}}\,\left(e^{2\lambda_{3,4}L}-1\right)\left(1+\left\langle \widehat{n}_{f2}\right\rangle \right)
\end{multline}

\noindent from which it is clear that the fluctuations in the system would be the result of the combined action of quantum and thermal noise, the latter contributed by the linear lossy waveguide. In the limit case of waveguides at zero temperature $T=0$, where no thermal photons populate the waveguides and therefore their expectation value vanish $\left\langle \widehat{n}_{f2}\right\rangle =0$, the variance reduce to $\left(\Delta\varPsi_{3,4}(L)\right)_{T0}^{2}=\frac{1}{4}e^{2\lambda_{3,4}L}+\frac{1}{\left(1+A_{3,4}^{2}\right)}\frac{\alpha}{4\lambda_{3,4}}\,\left(e^{2\lambda_{3,4}L}-1\right)$. Again, in the uncoupled waveguides limit $\kappa\rightarrow0$, where $\lambda_{3,4}\rightarrow\pm\alpha$, we recover the signature variances: i) $\left(\Delta\varPsi_{3}(L)\right)^{2}=\frac{1}{4}e^{2\alpha L}$ for the waveguide with parametric gain and ii) $\left(\Delta\varPsi_{4}(L)\right)^{2}=\frac{1}{4}$ for a linear lossy waveguide. Figure \ref{fig:Variance_vsCoup2Loss} (c) confirms that at $T=0$ none of the Hermitian eigenoperators exhibits a variance squeezed below the vacuum limit $\frac{1}{4}$, independently of the coupling-to-loss (gain) ratio. As expected, the maximum and minimum fluctuations corresponds to the uncoupled case.

These values are modified by the coupling strength between the waveguides until they balance at the corresponding phase-transition point, i.e., $\left(\Delta\varPsi_{3,4}(L)\right)^{2}$=$\frac{1}{4}+\frac{1}{4}\alpha L$. Similar to the linear loss-linear gain case, it is found that the scaling of the noise variance is linear with the length of the waveguide $\alpha L$, which is a much slower scaling that the exponential trend of the uncoupled waveguides. Therefore, this property is found to be insensitive to the specific gain/loss model. Moreover, at the phase transition point, it is found again that such linear scaling correspond to the first-order Taylor series approximation of the geometric mean
of the variances for the uncoupled waveguides. Finally, we note that this value is smaller than the linear gain/linear loss system, demonstrating that parametric amplification can compensate linear loss with a smaller noise production, albeit at the cost of being phase sensitive.

In the presence of thermal photons, as depicted in Figure \ref{fig:Variance_vsCoup2Loss_1nf} (c) for the case of one thermal photon $\left\langle \widehat{n}_{f2}\right\rangle =1$ in the system, we observe that one of the modes is more sensitive to the change, increasing its variance, while the other is almost unaffected, still the overall behavior is similar to the already described for $T=0$.

As for the variance dependence on the waveguide length (see Figure \ref{fig:Variance_vsL}), no squeezing is observed independently of the value of $L$. In the limit $L\to0$, we recover the characteristic vacuum fluctuations $\left(\Delta\widehat{\varPsi}_{n}(0)\right)_{T0}^{2}=\frac{1}{4}$. Therefore, for the balanced system squeezing can not be measured, although it is possible in non-balanced configurations where the eigenoperators associated to the other pair of eigenvalues are allowed to take real values.

\subsection{Linear gain-Parametric loss}

Finally, we study the remaining configuration combining linear and parametric gain and loss models, depicted in Figure \ref{fig:Wvg_sketch} (d). The system consists of a linear amplifying waveguide with gain $g$, coupled with strength $\kappa$ to a lossy waveguide with parametric loss modeled through a squeezing transformation with squeezing parameter $\xi$, as described by the coupled evolution equations:
\begin{equation}
\partial_{z}\widehat{a}_{1}\left(z\right)=g\widehat{a}_{1}\left(z\right)+i\kappa\widehat{a}_{2}\left(z\right)+\sqrt{2g}\widehat{f}_{1}^{\dagger}\left(z\right)
\end{equation}
\begin{equation}
\partial_{z}\widehat{a}_{2}\left(z\right)=\xi\widehat{a}_{2}^{\dagger}\left(z\right)+i\kappa\widehat{a}_{1}\left(z\right)
\end{equation}

\noindent or, equivalently
\begin{equation}
\mathbf{P}=\left[\begin{array}{cc}
g & i\kappa\\
i\kappa & 0
\end{array}\right]\quad\mathbf{Q}=\left[\begin{array}{cc}
0 & 0\\
0 & \xi
\end{array}\right]
\end{equation}

The eigenvalues for the specific case of balanced gain and loss $g=\xi$ are given by
\begin{equation}
\lambda_{1,2}=g\pm i\kappa
\end{equation}
\begin{equation}
\lambda_{3,4}=\pm\sqrt{g^{2}-\kappa^{2}}
\end{equation}

Comparing with the eigenvalues in the combined parametric gain - linear loss system, we observe a clear analogy, where $g$ plays the role of $-\alpha$ in (\ref{eq:NonGLinL_eigenval}). Therefore, a similar analysis and conclusions apply. For example, balanced gain and loss $g=\xi$ forbids the occurrence of real-valued $\lambda_{1,2}$ except for uncoupled waveguides, a situation that is not relevant to this study. Also, the pair of eigenvalues $\lambda_{3,4}$ exhibits a phase transition point in parameter space at $\kappa/g=1$, changing from real to complex values, as depicted in Figure \ref{fig:Eigenvalues} (d).

\subsubsection{Eigenvectors and eigenoperators for real eigenvalues}

The associated eigenvectors in the $g\geq\kappa$ regime are given by
\begin{equation}
\boldsymbol{v}_{3,4}=\frac{1}{\sqrt{A_{3,4}^{2}+1}}\begin{bmatrix}iA_{3,4},1,-iA_{3,4},1\end{bmatrix}^{T}\label{eq:LinGNonL_eigenvect}
\end{equation}

\noindent where $A_{3,4}=\frac{\kappa}{-g\pm\sqrt{g^{2}-\kappa^{2}}}$. Again, at the eigenvalues' phase-transition point the eigenvectors coalesce, with $A_{3}=A_{4}$. The Hermitian eigenoperators derived from the eigenvectors in (\ref{eq:LinGNonL_eigenvect}) can be written as
\begin{multline}
\widehat{\varPsi}_{3,4}\left(L\right)=\frac{1}{2}\left(\widehat{a}_{3,4}(0)+\widehat{a}_{3,4}^{\dagger}(0)\right)e^{\lambda_{n}L}+\\
\frac{\sqrt{2g}}{2\sqrt{A_{3,4}^{2}+1}}e^{\lambda_{3,4}L}\int_{0}^{L}dz'e^{-\lambda_{3,4}z'}\left(A_{3,4}\widehat{f}_{1}(z')+A_{3,4}^{*}\widehat{f}_{1}^{\dagger}(z')\right)
\end{multline}

\noindent where $\widehat{a}_{3,4}(0)=\frac{iA_{3,4}\widehat{a}_{1}(0)+\widehat{a}_{2}(0)}{\sqrt{1+A_{3,4}^{2}}}$.

\subsubsection{Eigenoperators variance}

The variance that results from the real eigenvalues can be computed as follows:
\begin{multline}
\left(\Delta\varPsi_{3,4}(L)\right)_{T}^{2}=\frac{1}{4}e^{2\lambda_{3,4}L}+\\
\frac{A_{3,4}^{2}}{1+A_{3,4}^{2}}\frac{g}{4\lambda_{n}}\left(e^{2\lambda_{3,4}L}-1\right)\left(1+\left\langle \widehat{n}_{f1}\right\rangle \right)\label{eq:LinGNonL_var}
\end{multline}

\noindent where as in all the previous configurations with linear processes involved, the fluctuations in the system result from the combined action of quantum and thermal noise. As expected, the thermal noise contribution originates from the linear amplifying waveguide. In the absence of thermal photons, i.e., $\left\langle \widehat{n}_{f1}\right\rangle =0$ for $T=0$, the variance reduce to $\left(\Delta\widehat{\varPsi}_{3,4}(L)\right)_{T0}^{2}=\frac{1}{4}e^{2\lambda_{3,4}L}+\frac{A_{3,4}^{2}}{\left(1+A_{3,4}^{2}\right)}\frac{g}{4\lambda_{3,4}}\left(e^{2\lambda_{3,4}L}-1\right)$. From Figure \ref{fig:Variance_vsCoup2Loss} (d) we can observe that at $T=0$ one of the eigenmodes exhibits fluctuations squeezed below the vacuum fluctuations limit. These results can be explained considering the maximum and minimum variances that can be obtained, and that corresponds to the limit $\kappa\rightarrow0$ of uncoupled waveguides, where $\lambda_{3,4}\rightarrow\pm g$. The maximum value of the fluctuations corresponds to $\left(\Delta\varPsi_{3}(L)\right)^{2}=\frac{1}{4}\left(2e^{2gL}-1\right)$ characteristic of waveguide with linear gain, while the minimum is $\left(\Delta\varPsi_{X,Y+}(L)\right)^{2}=\frac{1}{4}e^{-2gL}$ signature
of the squeezed quadrature in a waveguide with parametric loss. 

The most obvious difference with a system where both gain and losses are described through squeezing transformation is that squeezing is not preserved in the whole region where the eigenvalues are real. Therefore, at the phase transition point, where squeezed and anti-squeezed variances converge, we have that $\left(\Delta\varPsi_{3,4}(L)\right)^{2}$=$\frac{1}{4}+\frac{1}{4}\alpha L$ and the fluctuations are above the quantum vacuum limit $1/4$. This can be understood by considering that to obtain a target amplification
level, linear gain introduces more noise than its parametric counterpart; therefore, at the compensation point, the fluctuations for the configuration under study exceed those for coupled waveguides featuring parametric gain and loss. As in the previous configuration, at the compensation point, the scaling of the fluctuations is linear with $\alpha L$,
with the associated potential benefits. Again, we observe that gain and loss mechanisms of different nature can compensate each other. Also, at the compensation point, coinciding with the phase transition, the variances equal the geometric mean of those at the $\kappa\rightarrow0$ limit for sufficiently small values of $\alpha L$, showing that these
conclusions are insensitive with respect to the model of gain/loss. It is important to highlight that, different from the parametric gain - linear loss case, we do get squeezed fluctuations for the system with balanced gain and losses. For thermal population, as depicted in Figure \ref{fig:Variance_vsCoup2Loss_1nf} (d) for the specific case of one thermal photon in the system, i.e. $\left\langle \widehat{n}_{f1}\right\rangle =1$, we observe that the coupling-to-gain (loss) ratio $\kappa/g$ where the squeezed mode crosses the vacuum fluctuation limit is slightly reduced. The influence on the anti-squeezed variance is more evident, considering its value in the limit of infinitesimal waveguides.

Analyzed in terms of its dependence on the waveguide length $L$ (see Figure \ref{fig:Variance_vsL} (d)), we observe that for every value of $L$ the squeezed fluctuations in one mode prevails, exhibiting and exponential decay as predicted by the analytical result in (\ref{eq:LinGNonL_var}) and being in theory infinitesimal for sufficiently large waveguides. As expected, the other mode showcases an exponential scaling in the fluctuations. Again, in the limit $L\to0$, we recover the characteristic vacuum fluctuations $\left(\Delta\varPsi_{3,4}(0)\right)_{T0}^{2}=\frac{1}{4}$.

\section{Conclusions}

We discussed how different gain and loss mechanisms influence the noise produced in gain-loss compensated coupled photonic waveguides. Our results highlight universal properties independent of the gain and loss model, such as a phase transition in the eigenvalue structure, where all eigenvalues vanish and eigenvectors only coalesce in pairs. 
Simultaneously, some particularities arise. For instance, when only linear effects are present, the phase transition point corresponds to an exceptional point (EP) in a classical Non-Hermitian (NH) system. However, a quantum treatment prevents the coalescence of all eigenvectors as the eigenspace dimension doubles and noise terms are included. On the other hand, when only parametric phenomena are considered, the system's phase sensitivity impedes the existence of EPs. Our analysis also reveals that gain and loss mechanisms of
different nature can compensate each other, inheriting the phase sensitivity of the fluctuations in the parametric process and leading to squeezed fluctuations when the parametric loss is compensated through linear amplification. When linear phenomena are involved, the fluctuations result from combined thermal and quantum noise, unlike when only parametric
effects are present, and thus, the noise generated is exclusively quantum and corresponds to unamplified vacuum fluctuations. The gain-loss compensation and the phase transition points coincide for every analyzed waveguide configuration. Interestingly, although noise is unavoidably generated at these phase transition points, its scaling is linear with the waveguide length and outperforms the exponential scaling exhibited by uncoupled waveguides.

We believe these results contribute to deepening our understanding of NH quantum systems and the nontrivial properties at their phase transition points, with impact on fundamental research and potential applications in gain-loss compensation systems and electrically large photonic networks with reduced fluctuations.

\begin{acknowledgments}
I.L. acknowledges support from Ram\'on y Cajal fellowship RYC2018-024123-I and ERC Starting
Grant No. ERC-2020-STG948504-NZINATECH.
\end{acknowledgments}

\bibliography{apssamp_NH}

\end{document}